\documentclass[aps,prb,twocolumn,superscriptaddress,showpacs,floatfix]{revtex4}
\usepackage{graphicx}
\usepackage{epsfig}
\usepackage{amsmath}
\usepackage{amssymb}
\usepackage{subfigure}
\usepackage{mathptmx, microtype}

\begin{document}

\frenchspacing

\title{Temperature-dependent pressure-induced softening in Zn(CN)$_{2}$}

\author{Hong Fang}
\affiliation{Department of Earth Sciences, University of Cambridge, Downing Street, Cambridge, CB2 3EQ, UK}

\author{Anthony E. Phillips}
\affiliation{Centre for Condensed Matter and Materials Physics, School of Physics and Astronomy, Queen Mary University of London, Mile End Road, London, E1 4NS, UK}
\affiliation{Materials Research Institute, Queen Mary University of London, Mile End Road, London, E1 4NS, UK}

\author{Martin T. Dove}
\email{martin.dove@qmul.ac.uk}
\affiliation{Department of Earth Sciences, University of Cambridge, Downing Street, Cambridge, CB2 3EQ, UK}
\affiliation{Centre for Condensed Matter and Materials Physics, School of Physics and Astronomy, Queen Mary University of London, Mile End Road, London, E1 4NS, UK}
\affiliation{Materials Research Institute, Queen Mary University of London, Mile End Road, London, E1 4NS, UK}

\author{Matthew G. Tucker}
\affiliation{ISIS Facility, Rutherford Appleton Laboratory, Chilton, Didcot, Oxfordshire, OX11 0QX, UK}
\affiliation{Diamond Light Source, Harwell Campus, Didcot, Oxfordshire, OX11 0DE, UK}

\author{Andrew L. Goodwin}
\affiliation{Department of Chemistry, University of Oxford, Inorganic Chemistry Laboratory, South Parks Road, Oxford, OX1 3QR, UK}

\date{\today}
\begin{abstract}
We study the temperature dependence of the pressure-induced softening in the negative thermal expansion material Zn(CN)$_2$ using neutron powder diffraction and molecular dynamics simulations. Both the simulation and experiment show that the pressure-induced softening only occurs above a minimum temperature and also weakens at high temperatures. This is the first observation of the effects of temperature on pressure-induced softening.
\end{abstract}


\maketitle

\section{Introduction}

The phenomenon of pressure-induced softening---whereby the elastic stiffness of a material actually decreases on compression---is a counterintuitive mechanical response that is rarely observed in materials, at least over extended pressure ranges. The few known examples from experiment include silica glass,~\cite{Tsiok 1998} and two negative thermal expansion (NTE) framework materials ZrW$_2$O$_8$~\cite{Pantea 2006} and Zn(CN)$_2$.~\cite{Chapman 2007} We have recently predicted the existence of pressure-induced softening in a number of cubic zeolites, all of which shown NTE.~\cite{Fang zeolite}

The intuition that materials should become stiffer under compression is grounded in the observation that shortening bonds increases the steepness of the corresponding potential energy surface. Thus there is a strong fundamental physics interest in understanding the apparent violation of this basic mechanical response. It is also likely that the existence of this counterintuitive response in materials that display the equally rare phenomenon of volume NTE is not coincidental.~\cite{Fang zeolite} That pressure-induced softening may signal the existence of a host of unusual thermodynamic responses (\emph{e.g.}\ NTE) with important technological applications acts as further motivation to understand both its microscopic origin and its implications for other material properties.

In high-symmetry structures, elastic stiffness can be characterized by the bulk modulus, $B=-(\partial \ln V/\partial p)^{-1}_T$, where $V$ is the volume, $p$ is the pressure, and $T$ is the temperature.~\cite{Nye 1957} Typical values of $B_0=B(p=0)$ lie in the range 30--100\,GPa, with larger values corresponding to materials of increasing mechanical stiffness.~\cite{Newnham 2005} Experimentally, $B_0$ is usually measured by fitting the pressure dependence of the crystallographically-determined unit cell volume to an appropriate equation of state.~\cite{BM} The pressure dependence of the bulk modulus is characterized by the dimensionless parameter $B_0^\prime=(\partial B/\partial p)_T\mid_{p=0}$, which for many materials has a value in the vicinity of $+4$ at ambient temperature; indeed this is the value to which the widely-used second-order Birch-Murnaghan equation of state corresponds.~\cite{BM} So, in other words, for most materials one expects an increase of 5--10\% in bulk modulus for each 1\,GPa increase in hydrostatic pressure.

In this context, measured \textit{negative} values of $B_0^\prime = -17$ ($B_0=76$\,GPa)~\cite{Pantea 2006} for ZrW$_2$O$_8$, and $B_0^\prime = -6.0(7)$ ($B_0=34.19(21)$\,GPa)~\cite{Chapman 2007} or $B_0^\prime = -8.6(13)$ ($B_0=36.9(22)$\,GPa)~\cite{Collings 2013} for Zn(CN)$_2$ are strikingly anomalous. In both cases an applied pressure of 1\,GPa causes the material stiffness to \textit{decrease} by $\sim20$\%. Discussion of the origin of pressure-induced softening have been given in references~\onlinecite{Pantea 2006}, \onlinecite{He 2010} and \onlinecite{Walker 2007}, the first two corresponding in practice to a model that represents a two-dimensional idealization of Zn(CN)$_2$ (see Appendix~\ref{App_Pantea}). These works all associated the negative value of $B_0^\prime$ with dynamical effects associated with low-frequency phonon modes involving rotations of quasi-rigid polyhedral groups of atoms.

What has not yet been seen at all is the dependence of pressure-induced softening on temperature in any material. In this paper we study this effect in Zn(CN)$_2$ using both molecular dynamics (MD) simulations and variable-$p/T$ neutron powder diffraction. Our MD simulations reveal a very strong temperature dependence to $B_0^\prime$, which even involves a change in sign. Based on an interpretation of the relevant fluctuations, we propose an empirical form for the $B_0^\prime(T)$ function. A simple single-particle Hamiltonian of a rigid-unit mode system similar to Zn(CN)$_2$ was analyzed in Ref.~\onlinecite{Pantea 2006}, but this gives a prediction for the temperature dependence of pressure-induced softening that differs from the results presented in this paper; we discuss this in Appendix~\ref{App_Pantea}.

This paper begins with a brief discussion of the lattice dynamics of Zn(CN)$_2$, from which we predict qualitatively the form of the $B_0^\prime(T)$ function. We proceed to describe first the MD simulations we have performed, which establish the temperature-dependence of $B_0^\prime$ for this material, and second the neutron diffraction experiments that provide some confirmation of the MD results. We conclude with a discussion of the implications of our study for computational and experimental investigations of pressure-induced softening in related systems.

\section{Lattice dynamics and a phenomenological model}\label{section:lattice_dynamics}

\begin{figure}[t]
\begin{center}
\includegraphics[width=8 cm]{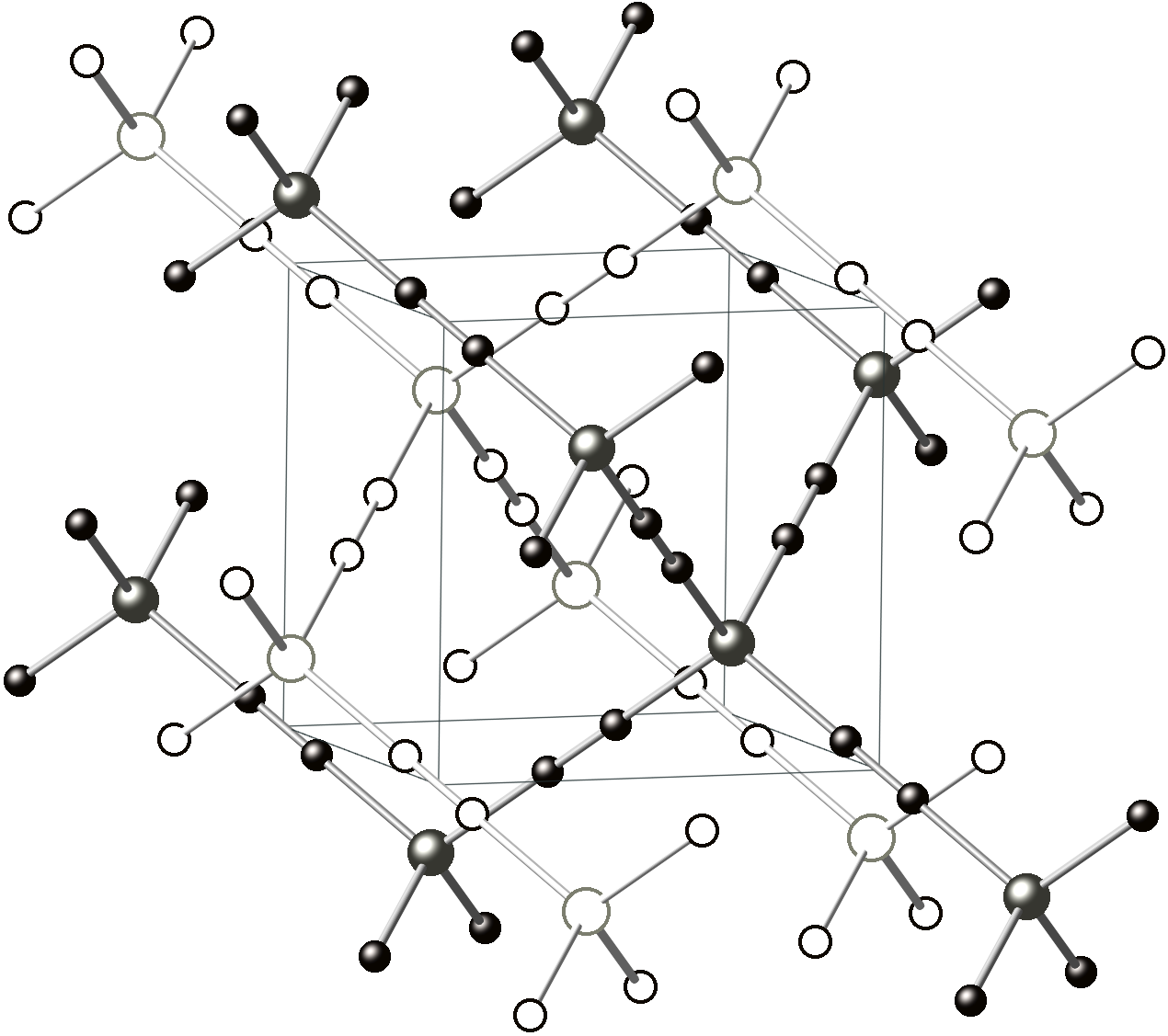}
\end{center}
\caption{\label{fig:structure} Crystal structure of Zn(CN)$_2$ showing the two sublattices where one is shown using photorealistic shading to represent the atoms and the other with the atoms represented by open spheres. The primary features are the tetrahedral coordination of the Zn atoms (larger spheres) to the CN molecular anions (smaller spheres connected by a short bond), and the linear Zn--cyanide--Zn connectivity.}
\end{figure}

From a lattice dynamics perspective, the key point of interest concerning zinc cyanide has always been its NTE behaviour:~\cite{Goodwin 2005} the volume of its cubic unit cell decreases on heating at a rate that is more than double that of longer-established NTE systems such as ZrW$_2$O$_8$.~\cite{Mary 1996} The crystal structure of Zn(CN)$_2$~\cite{Hoskins 1990} can be described as a network of zinc atoms connected via linear Zn--C--N--Zn linkages as shown in Fig.~\ref{fig:structure}. Each Zn centre is coordinated in a tetrahedral fashion by four C or N atoms.~\cite{Williams 1997} It is thought that, to a first approximation, these tetrahedral units are not readily deformed,~\cite{Goodwin 2005,Chapman 2005} so the dominant vibrational motion in zinc cyanide involves flexing of the Zn--C--N--Zn linkages as propagated in a family of low-energy transverse acoustic phonon modes with energies less than 1\,THz.~\cite{Chapman 2006,Zwanziger 2007,Mittal 2011,Fang 2013}

In the case of amorphous silica, molecular dynamics simulations suggested a relation between the bulk compressibility and the extent of network flexibility as measured by the magnitude of fluctuations involving whole-body rotations of SiO$_4$ tetrahedra.~\cite{Walker 2007} We can propose a similar explanation for the case of Zn(CN)$_2$, where at some finite temperature the structure is buckled through rotations and translations of the Zn(C/N)$_4$ tetrahedra and cyanide bridges---the same fluctuations that are responsible for NTE. An initial application of negative pressure to expand the structure will first be accommodated through straightening of the Zn--C--N--Zn linkages and alignment of the Zn(C/N)$_4$ tetrahedra, which costs relatively little energy. This process will then be followed later by stretching of the Zn--C--N--Zn linkages which costs rather more energy. Thus the bulk modulus is expected to increase with negative pressure, which implies that $\partial B_0 / \partial p$ takes a negative value.

\begin{figure}[t]
\begin{center}
\includegraphics{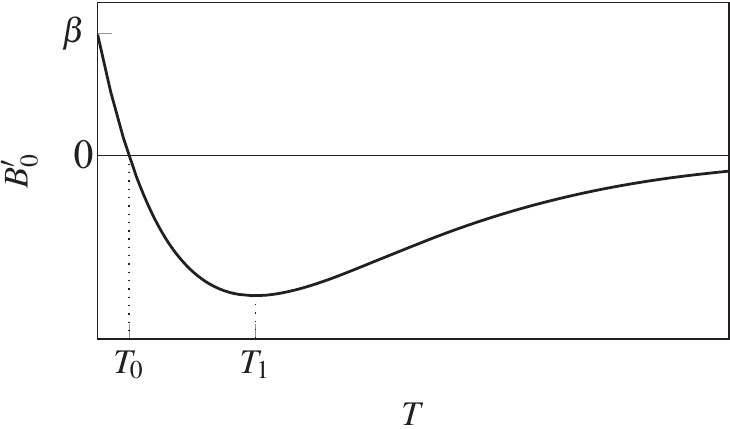}
\end{center}
\caption{\label{fig:qualcurve} Schematic curve showing the predicted variation of $B_0^\prime$ with temperature $T$ according to Eq.~\eqref{eq:qualitative}. This schematic defines the meanings of the parameters $\beta$, $T_0$ and $T_1$.}
\end{figure}

This explanation suggests that the magnitude of the pressure-induced softening should depend on temperature. At low temperature there will be no fluctuations to accommodate stretching of the structure, and we therefore expect a positive value of $B_0^\prime$. Indeed, without fluctuations, compression of the structure will also require compression of bonds. Thus we expect pressure-induced softening to occur only when there are thermal fluctuations, and we expect that on heating $B_0^\prime$ will initially have a positive value, which will reduce and become negative at some finite temperature. At higher temperatures we might expect not to see the transition between straightening the fluctuations and stretching of bonds, so that $B_0$ will become relatively constant with pressure and hence $B_0^\prime$ will tend towards zero from its negative value.

We model the variation of $B_0^\prime$ with temperature $T$ in Zn(CN)$_2$ using the following phenomenological form:
\begin{equation}
  \label{eq:qualitative}
  B_0^\prime = -\beta\left(\frac{T-T_0}{T_0}\right)\exp\left(-\frac{T}{T_1-T_0}\right),
\end{equation}
parameterised by the variables $\beta$, $T_0$, and $T_1$. This is illustrated in Fig.~\ref{fig:qualcurve}. We will find below that the temperature-dependence of $B_0^\prime$ in Zn(CN)$_2$ obtained from both molecular dynamics simulations and neutron powder diffraction experiments is described well by this phenomenological model.

\section{Molecular Dynamics Simulations}

\subsection{Method}

Molecular dynamics (MD) simulations were performed along isotherms of Zn(CN)$_2$ at selected temperatures with an interatomic potential model developed from \textit{ab initio} calculations.~\cite{Fang 2013} Atomic charges were obtained from a distributed multipole analysis.~\cite{DMA,camcasp} Morse potentials were used to describe the energies of chemical bonds, and bond angle terms were used to define the Zn(C/N)$_4$ tetrahedra. A linear three-body potential, $E_L = K(1-\cos \varphi)$, was used to describe the transverse vibrations of C and N in the linkages Zn--C(N)--N(C)--Zn. For these potentials, parameters were tuned by fitting to quantum mechanical calculations based on small clusters. The dispersion interactions between carbon and nitrogen atoms were modeled using the Buckingham potentials of reference \onlinecite{Williams}. Further details are given in the Supplementary Information and Ref.~\onlinecite{Fang 2013}; Ref.~\onlinecite{Fang 2013} also provides comparison of the performance of the model against experimental data.

The MD simulations were performed using the DL\_POLY code,~\cite{Todorov 2006} using a sample described as a $10\times10\times10$ supercell containing $10,000$ atoms. Normal periodic-boundary conditions were used. A constant-stress and constant-temperature ensemble with a Nos\'{e}-Hoover thermostat~\cite{Hoover 1985} was used. The equations of motion were integrated using the leapfrog algorithm with a time step of $0.001$ ps.

Each simulation was performed with an equilibration time of 20 ps followed by a run of 40 ps from which an average volume was calculated from writing the instantaneous volume at every 0.02 ps. Altogether we performed 242 MD simulations.

\begin{figure}[t]
\begin{center}
\includegraphics[width=8.5cm]{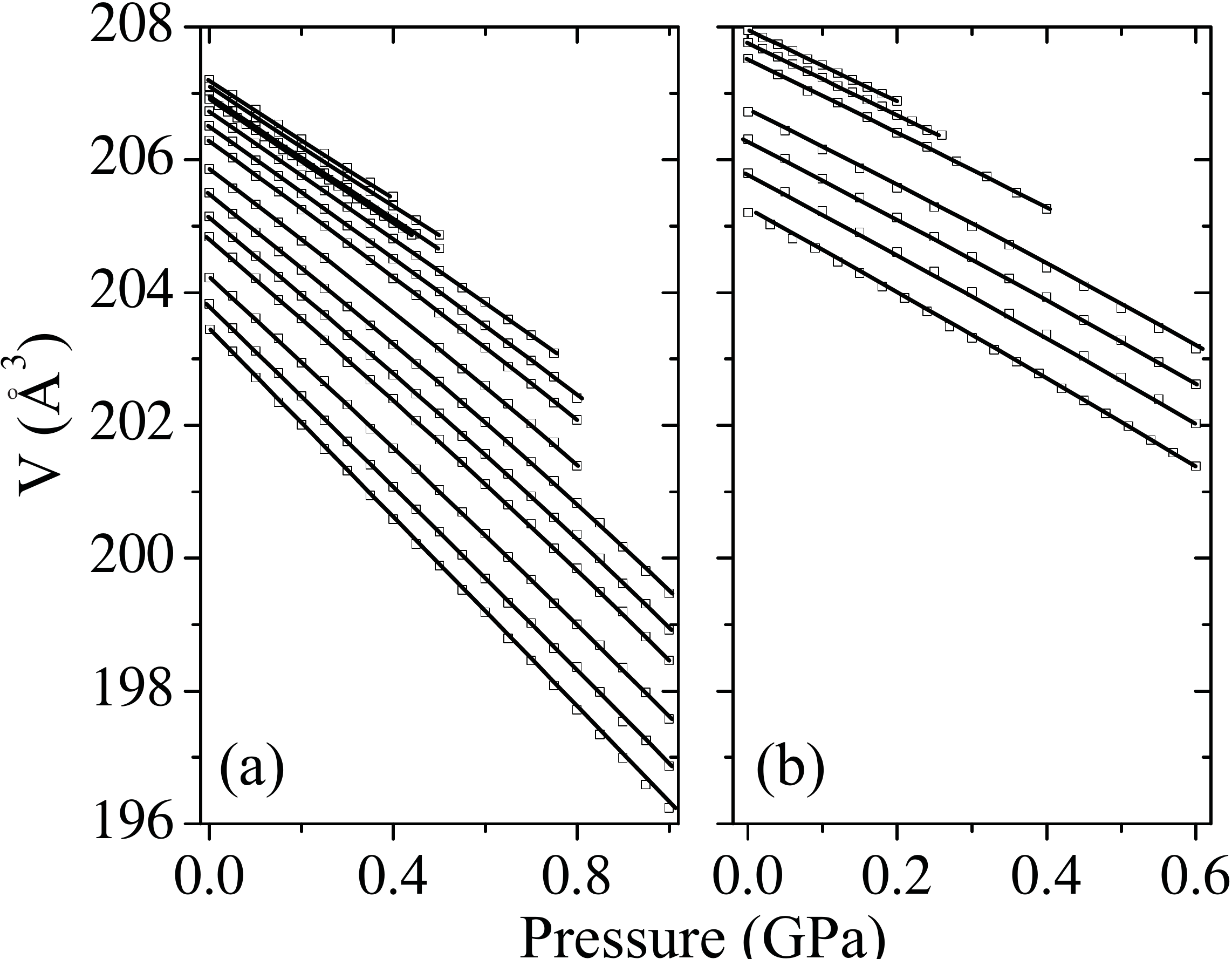}
\end{center}
\caption{\label{fig:overallfit} Simultaneous fit to: (a) the isotherm data sets (open squares) at (from up to down in the plot) 1, 10, 25, 30, 50, 75, 100, 150, 200, 250, 300, 400, 500 and 600\,K from the MD; (b) the isotherm data sets (open squares) at (from up to down in the plot) 30, 50, 75, 150, 200, 250 and 300\,K from the neutron diffraction, using the third-order Birch-Murnaghan equation of state. $V_0(T)$ and $B_0(T)$ in the equation of state were expressed as quadratic polynomials. $B_0^{\prime}(T)$ was put into the phenomenological form of Eq.~\ref{eq:qualitative}.}
\end{figure}

\subsection{Equation of state from simulation data}

\begin{figure}[t]
\begin{center}
\includegraphics[width=8cm]{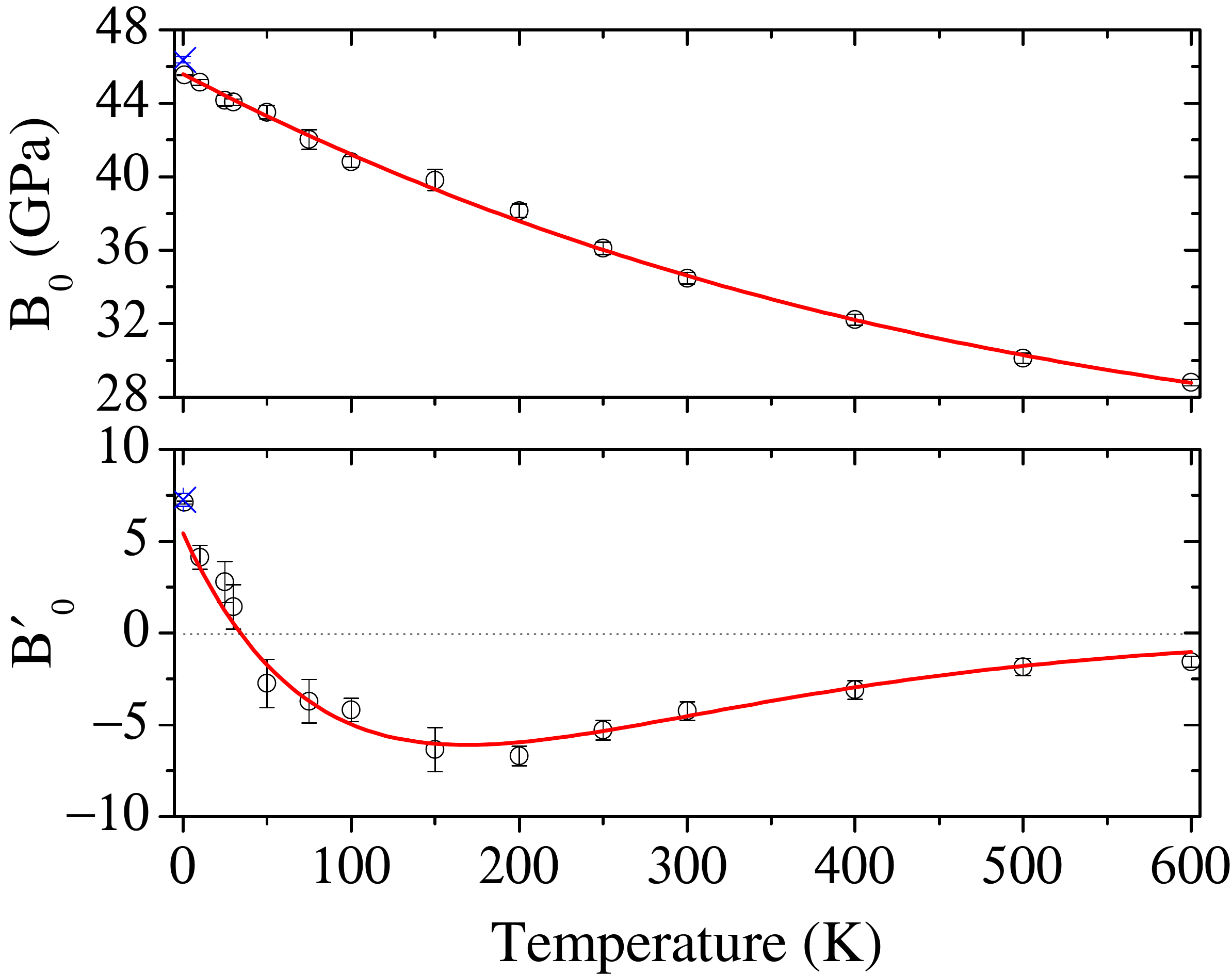}
\end{center}
\caption{\label{fig:MD} Bulk modulus at zero pressure $B_0$ and its first derivative $B_0^\prime$ as functions of temperature, obtained from fitting the third-order Birch-Murnaghan equation of state to the simulated isotherms of Zn(CN)$_2$. Values at zero temperature calculated in harmonic lattice dynamics using GULP~\cite{Gale 1997} are also included in the plot (blue cross). Red curves show the calculated $B_0(T)$ and $B_0^{\prime}(T)$ using values of the parameters from the simultaneous fit, where $B_0(T)$ was described by a quadratic polynomial and $B_0^{\prime}(T)$ was expressed by Eq.~\ref{eq:qualitative}.}
\end{figure}

Plots of sample volume as a function of pressure for the range of temperatures are given in Fig.~\ref{fig:overallfit}(a). Values of $B_0$ and its derivative $B_0^\prime$ for each temperature were obtained by fitting a third-order Birch-Murnaghan equation of state to the data, treating each temperature run separately, and shown in Fig.~\ref{fig:MD} and provided in Tab.~\ref{tab:results} for later comparison with experimental data.

\begin{table}[t]
\setlength{\tabcolsep}{4pt}
\caption{Simulation (Sim.) and neutron powder diffraction (Exp.) values of $B_0$ (units of GPa) and $B_0^\prime$ obtained as described in the text. Most of the simulation results are obtained using the MD method; the exceptions are the values at 0\,K, which are taken from a separate simulation of the equation of state using the lattice energy code GULP.~\cite{Gale 1997}}
\centering
\begin{tabular}{c| c| c| c| c}
  \hline\hline
  $T$ (K) & $B_0$ (Sim.) & $B_0$ (Exp.) &$B_0^\prime$ (Sim.) &$B_0^\prime$ (Exp.) \\ [0.5ex]
  \hline
  0  &46.37(17)& -- & $7.2(3)$ & --   \\ [0.5ex]
  1  &45.54(2)& -- & $7.1(1)$ & -- \\ [0.5ex]
  10 &45.15(17)& -- & $4.1(6)$  & --  \\ [0.5ex]
  25 &44.15(29)& -- & $3(1)$    & --  \\ [0.5ex]
  30 &44.07(16)& 39.1(7) &1(1) & $-2(7)$ \\ [0.5ex]
  50 &43.5(4)& 38.5(9) & $-3(1)$ & $-3(6)$ \\ [0.5ex]
  75 &42.0(5)& 38.8(10) & $-4(1)$ & $-8(5)$ \\ [0.5ex]
  100&40.8(3)& --         & $-4.2(6)$ & --        \\ [0.5ex]
  150&39.8(6)& 36.8(4)& $-6(1)$ & $-7(1)$ \\ [0.5ex]
  200&38.2(4)& 36.1(3)& $-6.7(5)$ & $-8.1(8)$ \\ [0.5ex]
  250&36.1(3)& 35.6(3)& $-5.3(5)$ & $-9.0(7)$  \\ [0.5ex]
  300&34.5(3)& 33.4(3)& $-4.2(5)$ & $-4(1)$  \\ [0.5ex]
  400&32.2(3)& --         & $-3.1(5)$ & --         \\ [0.5ex]
  500&30.11(28)& --         & $-1.8(5)$ & --         \\ [0.5ex]
  600&28.79(18)& --         & $-1.6(3)$ & --         \\ [0.5ex]
  \hline
\end{tabular}
\label{tab:results}
\end{table}

As a subsequent task, we refitted the data in Fig.~\ref{fig:overallfit}(a) by expressing $B_0^{\prime}$ by the phenomenological form of Eq.~\ref{eq:qualitative} and using quadratic polynomials to describe the zero-pressure volume $V_0(T)$ and the zero-pressure bulk modulus $B_0(T)$ as functions of temperature. In this case we performed an overall fit rather than fitting to each temperature data separately, adjusting the values of the parameters $\beta$, $T_0$ and $T_1$ in Eq.~\ref{eq:qualitative}, together with the parameters in the polynomials of volume and bulk modulus. Fig.~\ref{fig:overallfit}(a) and Fig.~\ref{fig:MD} show the good quality of the fit, and highlight the consistency of the phenomenological model with the MD data.

\begin{table}[t]
\setlength{\tabcolsep}{4pt}
\caption{Comparison of the values of $\alpha$ (MK$^{-1}$), $\partial B_0 /\partial T$ (GPa/K) ($T=300$\,K) and $\partial \alpha/\partial p$ (MK$^{-1}$GPa$^{-1}$) ($T=300$\,K) from the MD and experimental studies.}
\centering
\begin{tabular}{c| c| c| c}
  \hline\hline
   & $\alpha$ & $\partial B_0/\partial T$ & $\partial \alpha/\partial p$  \\ [0.5ex]
  \hline
  & $-48.9$ to $-40.9$ [10 to 100 K] \cite{Fang 2013} & & \\[-1ex]
  \raisebox{1.5ex}{MD} & $-30.1(4)$ [300 K] & \raisebox{1.5ex}{$-0.027(2)$} & \raisebox{1.5ex}{$-23(2)$} \\ [0.5ex]
  \hline
  Exp. & $-47.7(7)$ [averaged over 30 to 300 K] & $-0.035(23)$ & $-31(21)$  \\ [1ex]
  \hline
\end{tabular}
\label{tab:comparison}
\end{table}

The fitted values of $\beta$, $T_0$ and $T_1$ in Eq.~\ref{eq:qualitative} are 5.4(4), 34(2)\,K and 168(8)\,K, respectively. From the simultaneous fit, we also obtained the coefficient of thermal expansion $\alpha = \partial \ln V_0 / \partial T$ and the derivative $\partial B_0 / \partial T$; values for $T=300$\,K are given in Tab.~\ref{tab:comparison} for later comparison with the experimental results (below). According to the thermodynamic expressions for $\alpha$ and $B$ combined with the Maxwell relation $\partial^2 V / (\partial T \partial p)_{p,T}= \partial^2 V / (\partial p \partial T)_{T,p}$, the variation of $\alpha$ with pressure can be calculated as

\begin{equation}
  \label{eq:thermodynamic}
\left( {\frac{{\partial \alpha }}{{\partial p}}} \right)_T  = \frac{1}{{B^2 }}\left( {\frac{{\partial B}}{{\partial T}}} \right)_p
\end{equation}

\noindent The value of $\partial \alpha / \partial p$ at $p=0$ and $T=300$\,K is also given in Tab.~\ref{tab:comparison}.

In the limit of the temperature decreasing to zero, the value of $B_0^\prime$ from the MD simulations becomes positive. The zero-temperature value given in Tab.~\ref{tab:results} and Fig.~\ref{fig:MD} has been computed from a lattice energy calculation using the GULP code.~\cite{Gale 1997} It is clear from Fig.~\ref{fig:MD} that the MD results are tending towards the zero-temperature lattice energy result. The ambient-temperature values of $B_0$ and $B_0^\prime$ calculated by MD are consistent with those reported in previous experimental studies cited in the Introduction.~\cite{Chapman 2007,Collings 2013}

It is possible that the values of $B_0$ and $B_0^\prime$ obtained from fitting to the MD isotherms might not be directly comparable with experiment at low temperatures, because the MD simulations follow the classical equipartition of energy. However, for NTE materials such as Zn(CN)$_2$, most of the flexibility comes from vibrational modes that have low frequencies of the order of $\leq1$\,THz. These modes contribute most strongly to NTE through their large negative Gr\"{u}neisen parameters,~\cite{Fang 2013} which suggests that even at low temperatures such as $\sim 50$\,K, these modes remain populated and hence contribute to the dynamics of the material. This is born out from measurements of the variation of $\alpha$ with temperature,~\cite{Goodwin 2005} which show little sign of departure from the classical equipartition of energy down to very low temperatures. Another possible concern is that zero-point motion of the NTE modes is sufficient to allow a bond-bending deformation mechanism even in the limit $T\rightarrow0$\,K. Using harmonic lattice dynamics calculations performed using the GULP code~\cite{Gale 1997} we have estimated the zero-point contribution to $B_0^\prime$ at 0\,K to be $\simeq-0.1$ (as shown in Appendix~\ref{App_ZPM}), and hence a negligible correction to the values given in Fig.~\ref{fig:MD}. For these two reasons we do not anticipate much discrepancy between the classical MD results and the true quantum picture even at relatively low temperatures.

\section{Neutron Powder Diffraction}

\subsection{Method}

In order to confirm the predicted variation of $B_0$ and $B_0^{\prime}$ with temperature, we carried out a neutron powder diffraction experiment using the GEM diffractometer at ISIS.~\cite{Hannon 2005} A polycrystalline Zn(CN)$_2$ sample was contained within a Ti-Zr alloy pressure cell, which produces a featureless background in the diffraction pattern; this was itself contained within a closed-cycle refrigerator, allowing us to control both temperature and pressure. Hydrostatic pressure applied to the sample was generated by an external pressure intensifier unit to an accuracy of 1\,bar. The sample temperature was controlled by a closed cycle refrigerator (CCR) that can operate within a temperature range of 4 to 325\,K. Diffraction data were collected at various pressures from 5\,bar to a value limited by the phase diagram of helium gas at temperatures of 30, 50, 75, 150, 200, 250 and 300\,K~\cite{SI}. Measurements at lower temperatures could not be carried out because the helium gas would readily have liquefied at low pressure, making the true sample pressure unknown. For the same reasons, the accessible pressure range is much lower for the very low temperature data points. This limitation has the unfortunate consequence of increasing the experimental uncertainty in the derived values of $B_0$ and $B_0^\prime$ for temperatures below 100\,K.

\begin{figure}[t]
\begin{center}
\includegraphics[width=8cm]{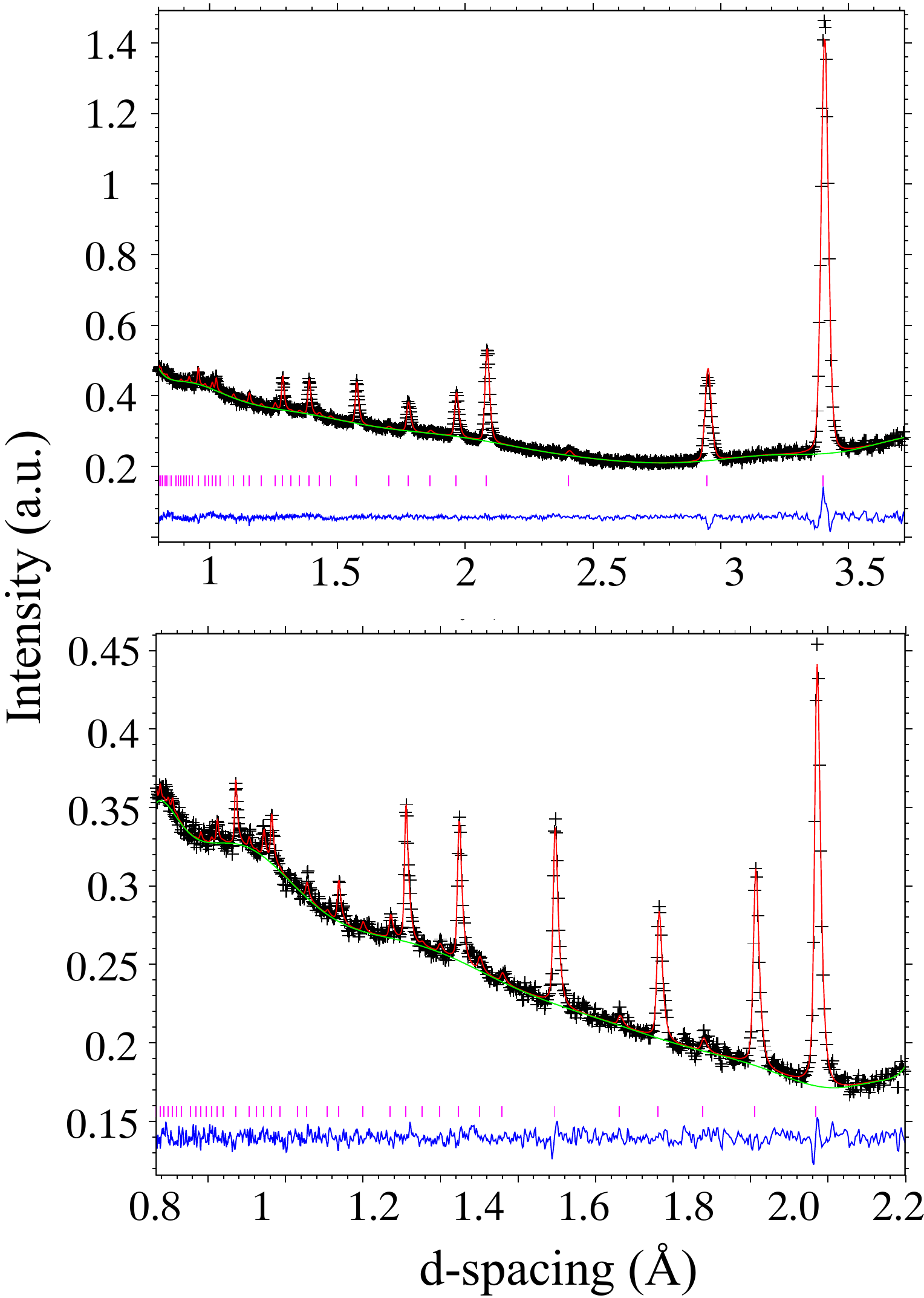}
\end{center}
\caption{\label{fig:Rietveld} A representative Rietveld fit to the neutron diffraction data collected using the GEM instrument at ISIS. These data correspond to a sample temperature of 300\,K and a hydrostatic pressure of 1.2\,kbar. Data points are shown as black crosses, the fit as a red line, the background function as a green line, and the residual (data$-$ fit) as a blue line.}
\end{figure}

Rietveld refinement of the neutron diffraction patterns using GSAS~\cite{Larson 2000} and EXPGUI~\cite{Toby 2001} yielded satisfactory fits, as shown in Fig.~\ref{fig:Rietveld}, and enabled determination of the unit cell volume to an accuracy of 0.001\,\AA$^3$. The structural model used was that described in Ref.~\onlinecite{Williams 1997}, which takes into account the head-to-tail disorder of the cyanide ions. The refined structural parameters were the position of the C/N atoms (constrained by the $Pn\bar{3}m$ crystal symmetry to a single parameter), the anisotropic atomic displacement parameters for the C/N atoms (constrained to be the same for C and N), and the Zn isotropic atomic displacement parameter. A table of refined structural parameters for all data sets is given in the Supplementary Information.~\cite{SI}

\subsection{Equation of state from experimental data}

\begin{figure}[t]
\begin{center}
\includegraphics[width=8cm]{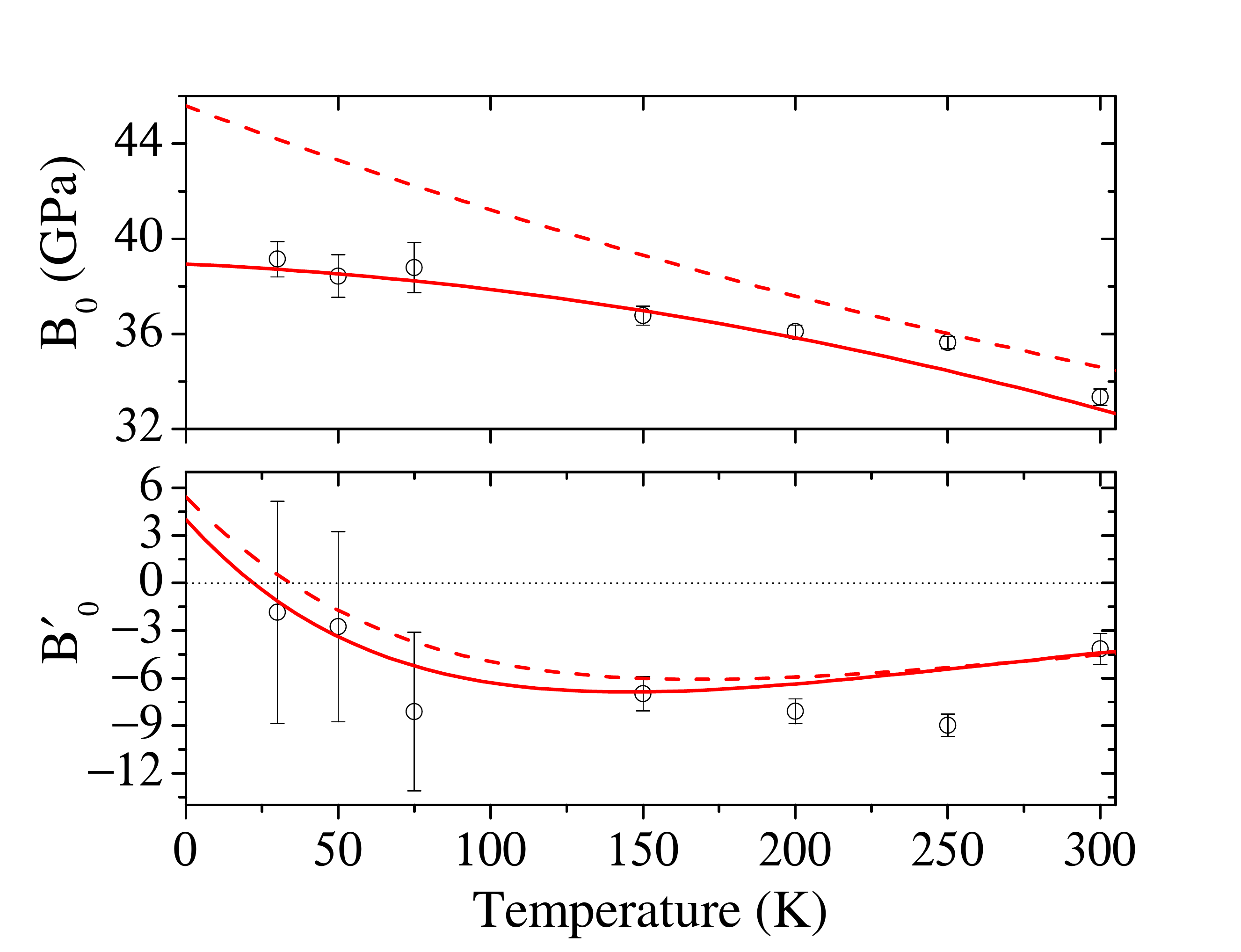}
\end{center}
\caption{\label{fig:Exp} Bulk modulus at zero pressure $B_0$ and its first derivative $B^\prime$ as functions of temperature, obtained from fitting the third-order Birch-Murnaghan equation of state to the experimental isotherms of Zn(CN)$_2$. Solid curves show the calculated $B_0(T)$ and $B_0^{\prime}(T)$ using values of the parameters from the simultaneous fit, where $B_0(T)$ was described by a quadratic polynomial and $B_0^{\prime}(T)$ was expressed by Eq.~\ref{eq:qualitative}. The relatively large deviation of the $B^\prime_0$ curve from the data points beyond 150 K is consistent with the relatively large fitting error for the corresponding isotherms in Fig.3(b). Dashed curves are the results from the MD for comparison (also seen in Table~\ref{tab:results}).}
\end{figure}

Values of $B_0$ and $B_0^\prime$ were extracted from the experimental lattice parameter data using the same approach as in the analysis of the MD results, fitting each isotherm to a third-order Birch-Murnaghan equation of states.~\cite{BM} The corresponding thermal evolution of derived values of $B_0$ and $B^\prime$ are shown in Fig.~\ref{fig:Exp}, and the numerical values of  $B_0$ and $B_0^\prime$ are listed in Table~\ref{tab:results} for comparison with the results from the MD simulations. We consider that the agreement between the experimental and MD values is reasonable given the difficulties in the experiment and the fact that the intermolecular potential was not tuned against experimental data. We will comment more on the level of agreement below.

We further carried out a simultaneous fit to all the $p$--$V$ data sets as we did for the MD data. The quality of the fitting is shown in Fig.~\ref{fig:overallfit}(b) and Fig.~\ref{fig:Exp}. The values of $\beta, T_0$ and $T_1$ in Eq.~\ref{eq:qualitative} obtained from this simultaneous fit are 4(5), 22(14)\,K and 145(29)\,K, respectively. These are in reasonable agreement with the values obtained by fitting to the MD data (Fig.~\ref{fig:MD}), albeit with rather larger values of the standard deviations on each parameter. The consistency between the MD and neutron diffraction results is highlighted by comparing the fitted curves for $B_0$ and $B_0^{\prime}(T)$ in Fig.~\ref{fig:Exp}. In the case of $B_0$, agreement is within $10\%$, although the curvature differs at low temperature. This may be a systematic error coming from the fact that at low temperature we had access to a much more restricted range of pressures. On the other hand, the agreement between the MD and neutron diffraction results for $B_0^{\prime}$ is much closer.

Values of $\alpha$ and $\partial B_0 / \partial T$ at 300\,K extracted from our data are given in Tab.~\ref{tab:comparison}, together with the value of $\partial \alpha / \partial p$ obtained from Eq.~\ref{eq:thermodynamic}. The value of $\alpha$ we have determined from experiment is consistent with the results reported in other experiments.~\cite{Goodwin 2005,Chapman 2007} Altogether, the experimental values given in Tab.~\ref{tab:comparison} are in fair agreement with the results from the MD simulations, when account is taken of the relatively large errors for the experimental data. We note that the value of $\alpha$ from experiment agrees well with the values from MD in the temperature range of 10--100\,K, but the value of $\alpha$ from the MD simulations decreases faster on heating than in the experiment, suggesting an overestimate of anharmonicity in our model at high temperatures. Despite the different low-temperature curvatures of the fitted $B_0(T)$, the MD values of $\partial B_0 / \partial T$ and $\partial \alpha / \partial p$ at 300\,K agree to the experiment within the error as shown in Table~\ref{tab:comparison}.

Whilst we have not been able to measure diffraction data at sufficiently low temperatures to observe a definitive transition from negative to positive values of $B_0^{\prime}$, so that Fig.~\ref{fig:Exp} cannot be said to confirm the detailed MD results exactly, the agreement between the fitted $B_0^{\prime}$ curve and the data points nevertheless demonstrates that our data remain consistent with the phenomenological model embodied by Eq.~\ref{eq:qualitative}.

\section{Conclusions}

We have proposed a phenomenological model of pressure-induced softening in zinc cyanide, based on the well-established difference in energy between vibrations that involve sideways buckling of the cyanide ions and those that involve bond stretching. The functional form of equation \ref{eq:qualitative}, chosen to illustrate this qualitative model, is consistent with both experimental and simulated data.

In this model, the parameter $T_0$ represents the temperature above which $B_0^\prime$ becomes negative, and as such it is a measure of the temperature at which the structure starts to crumple. In other words, it corresponds to the temperature at which the vibrational modes responsible for negative thermal expansion begin to be substantially occupied. Encouragingly, these modes are known to have an energy less than 1\,THz $= k_\mathrm{B}\times48$\,K,~\cite{Chapman 2006,Zwanziger 2007,Mittal 2011,Fang 2013} which agrees with the values of $T_0$ obtained from fits to the experimental and MD data.

The importance of revealing the temperature dependence of $B_0^\prime$ transcends the specific case of Zn(CN)$_2$, as important as this particular material is. We have suggested elsewhere~\cite{Fang zeolite} that a negative value of $B_0^\prime$ is a common feature of NTE materials, having shown through extensive simulations of many cubic zeolites that almost all exhibit both NTE and pressure-induced softening effects. Indeed the establishment here of a temperature dependence in which the parameter changes sign on cooling is crucial for the interpretation of simulation studies. We envisage that the striking prediction of negative values of $B_0^\prime$ in many NTE materials will stimulate a growing number of \textit{ab initio} studies, but we caution that without accounting for thermal fluctuations such calculations may predict incorrectly a positive value of $B_0^\prime$.

\appendix

\section{Some comments on the single-particle Hamiltonian of Reference 2}\label{App_Pantea}

\begin{figure}[t]
\begin{center}
\includegraphics[width=6cm]{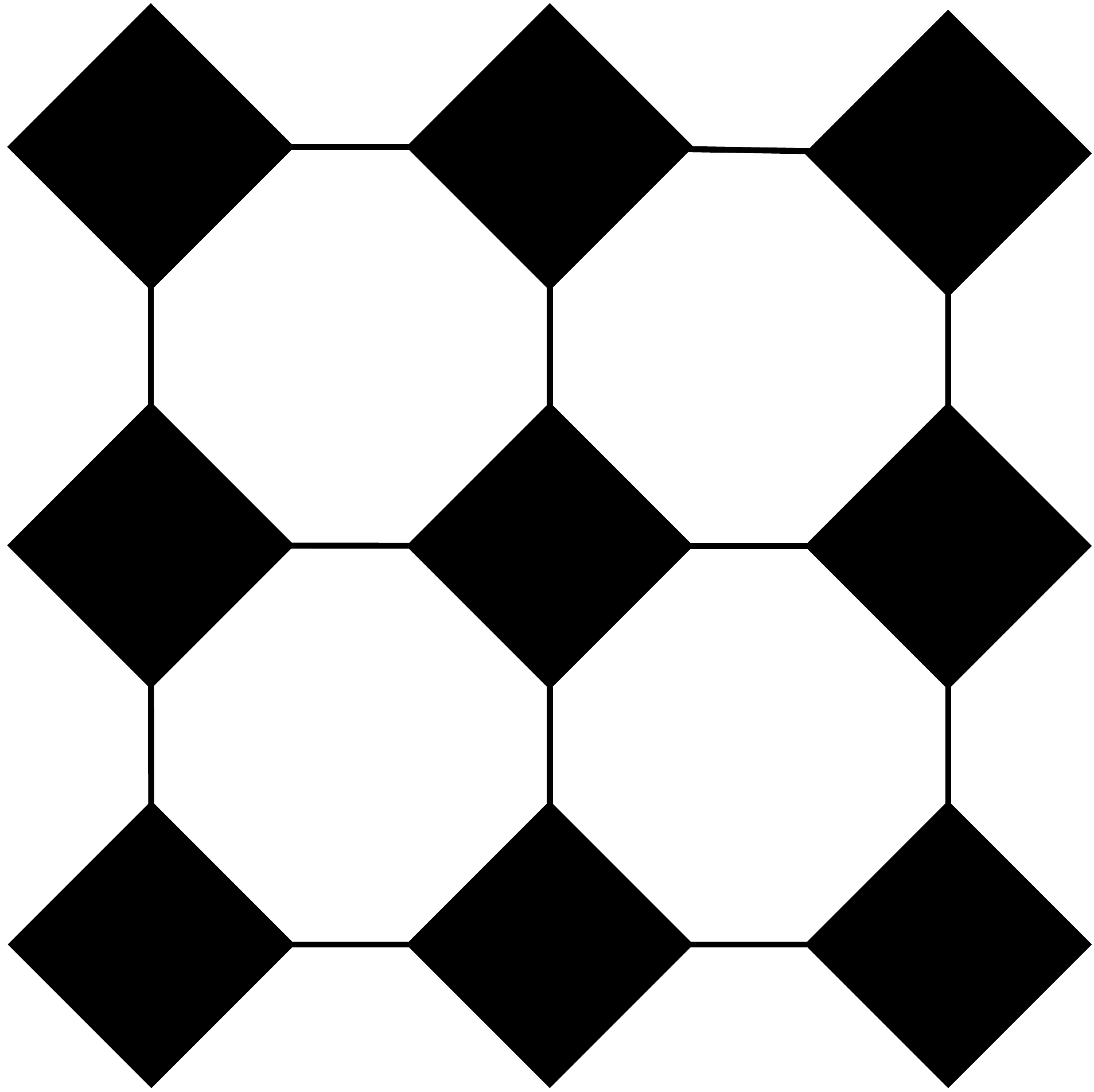}
\end{center}
\caption{\label{fig:2Dmodel} Representation of the two-dimensional model presented in Ref.~\onlinecite{Pantea 2006}, where we represented the square containing non-bridging vertex bonds of the original model by a rod that could represent the CN molecular ion of Zn(CN)$_2$. The model of Ref.~\onlinecite{Pantea 2006} has a single rotational variable for each square and, by extension, to each rod.}
\end{figure}

Ref.~\onlinecite{Pantea 2006} presents an experimental study of the pressure-dependence of the elasticity of ZrW$_2$O$_8$, showing pressure-induced softening. A simple model initially presented in Ref.~\onlinecite{Simon 2001} was used to provide an interpretation. The model is a two-dimensional array of corner-sharing squares, with one quarter of the squares missing, giving a structure with 3 polyhedra per unit cell of which 2 have have their bonds not connected to another square. This could be said to represent the existence of the non-bridging W--O bonds in ZrW$_2$O$_8$. However, the squares with 2 non-bridging bonds could be replaced by a rod connecting the fully-connected squares with no changes to the model, as developed in Ref.~\onlinecite{He 2010} and shown in Fig.~\ref{fig:2Dmodel}. As such, the model is more accurately a two-dimensional representation of Zn(CN)$_2$, albeit with some significant differences.~\cite{2D differences}

In the analysis of Ref.~\onlinecite{Pantea 2006}, a model Hamiltonian was constructed. This involves a single variable per square, namely the rotation angle $\theta_i$, and involves two energy terms. The first is a $pV$-like term,~\cite{not really}

\begin{equation}
E_{pV} = - p \left( 1 - \cos \theta_i \right)
\end{equation}

\noindent where the volume $V$ is reduced by rotations of the squares. This term automatically lowers the energy for rotations of the square, since these lead to a reduction in the volume. The second term is an anharmonic restoring force in even powers of $\theta_i$,

\begin{equation}
E_\mathrm{rot} = \frac{1}{2} K \theta_i^2 + \frac{1}{4} \gamma \theta_i^4 + \frac{1}{6} \delta \theta_i^6 + \cdots
\end{equation}

\noindent and gives a rise in energy for rotations of the squares.

Because there is no coupling between the squares, the phonons will all have frequencies that are independent of wave vector. This is a reasonable approximation for a system in which all phonons are rigid unit modes~\cite{rums}. However, in the case of Zn(CN)$_2$ there is a significant coupling term that gives a dependence of frequency on wave vector, which in the two-dimensional model would mean that counter-rotations of nearest-neighbour squares (modes whose wave vectors are at the edge of the two-dimensional Brillouin zone) will have a much lower frequency than the zone centre modes where all squares rotate in the same direction.

At temperature $T = 0$ there are two states differentiated by the value of $p$, namely the state at low $p$ in which the energy is dominated by $E_\mathrm{rot}$ leading to the preference for $\theta_i = 0$ for all values of $i$, and the state for $p$ above a critical value in which the energy is dominated by $E_{pV}$ leading to non-zero values of $\theta_i$ and a lowering of the volume in equilibrium. The transition between these two states is increasingly blurred at high temperature. The variation of volume around the critical pressure gives a variation in the bulk modulus that changes with pressure. This is not a real phase transition because the Hamiltonian doesn't contain terms that couple the rotations of neighbouring squares.

From Fig.~5 in Ref.~\onlinecite{Pantea 2006} we can see (although this is not articulated in Ref.~\onlinecite{Pantea 2006}) that this model predicts $B_0^\prime = 0$ at $T = 0$, with a value of $B_0^\prime$ that becomes negative and increasingly so with higher temperature. This behaviour at $p = 0$ arises because of the blurring of the transition at higher temperature. Where the model differs from the result presented in this paper is that we observe $B_0^\prime > 0$ at low $T$, only becoming negative for temperatures above a particular temperature. This difference arises from the neglect in Ref.~\onlinecite{Pantea 2006} of the stiffness of the squares except in an unspecified \textit{post hoc} manner. Thus to become consistent with the present study, the model should be extended by allowing an explicit non-infinite stiffness of the squares. This would give a finite compressibility at zero temperature, leading to a positive value of  $B_0^\prime$. We have shown separately (to be submitted) that the model described by Fig.~\ref{fig:2Dmodel} with a finite compressibility of the squares does indeed lead to a variation of $B_0^\prime$ with temperature that closely follows the phenomenological form of Eq.~\ref{eq:qualitative}.

\section{Effect of zero point motion}\label{App_ZPM}

Here we calculate the effect of zero-point energy on the value of $B_0^{\prime}$, which was not included in the lattice dynamics or the MD simulations that gave a positive value of $B_0^{\prime}$ at $T=0$. Thermodynamic calculations show that the contribution from the zero-point energy term to the value of $B_0^{\prime}$ of Zn(CN)$_2$ is a negligibly small negative number.

To derive $B_0^{\prime}$, we start from the pressure of an insulating crystal~\cite{Ashcroft 1976}

\begin{eqnarray}\label{eq1}
p =  - \frac{{\partial \Phi }}{{\partial V}} + \sum\limits_s {\left( {\frac{{\hbar \omega _s }}{V}\gamma _s n_s } \right)}
\end{eqnarray}

\noindent where $\Phi$ is the lattice energy of the crystal at zero temperature. The sum is over all the phonon modes $s={j,\textbf{k}}$ in the system with the angular frequency $\omega_s$ and Gr{\"u}neisen parameter $\gamma_s$. The average phonon occupation number of each mode is

\begin{eqnarray}\label{eq2}
n_s  = \frac{1}{{\exp (\hbar \omega /\tau ) - 1}} + \frac{1}{2}
\end{eqnarray}

\noindent with $\tau=k_\mathrm{B}T$ the temperature in units of energy. At zero temperature, $n_s=1/2$ and

\begin{eqnarray}\label{eq3}
 p|_{\tau  = 0}  &=&  - \frac{{\partial \Phi }}{{\partial V}} + \frac{1}{2}\sum\limits_s {\frac{{\hbar \omega _s }}{V}\gamma _s }  \nonumber \\
 & =&  - \frac{{\partial \Phi }}{{\partial V}} + \pi_0
\end{eqnarray}

\noindent where

\begin{eqnarray}\label{eq4}
\pi_0 = \frac{1}{2}\sum\limits_s {\frac{{\hbar \omega _s }}{V}\gamma _s }
\end{eqnarray}

\noindent is the contribution from zero-point energy. According to thermodynamic relations, the bulk modulus at zero temperature is

\begin{eqnarray}\label{eq5}
B|_{\tau  = 0}  =  - V\frac{{\partial p|_{\tau  = 0} }}{{\partial V}} = V\frac{{\partial ^2 \Phi }}{{\partial V^2 }} - \frac{{\partial \pi_0}}{{\partial \ln V}}
\end{eqnarray}

\noindent and the first derivative of bulk modulus at zero temperature is,

\begin{eqnarray}\label{eq6}
 B^{\prime}|_{\tau  = 0}  &=& \frac{{\partial B|_{\tau  = 0} }}{{\partial p}} \nonumber \\
  &=&  - \frac{V}{{B|_{\tau  = 0} }}\left[ {\frac{{\partial ^2 \Phi }}{{\partial V^2 }} + V\frac{{\partial ^3 \Phi }}{{\partial V^3 }}} \right] + \frac{V}{{B|_{\tau  = 0} }}\frac{\partial }{{\partial V}}\left( {\frac{{\partial \pi_0}}{{\partial \ln V}}} \right) \nonumber \\ [1.5ex]
  &=& B^{\prime}|_{\tau  = 0}^{LD}  + B^{\prime}|_{\tau  = 0}^{ZP}
\end{eqnarray}

\noindent with $V$ the volume of the crystal. In Eq.~\ref{eq6},

\begin{eqnarray}\label{eq7}
B^{\prime}|_{\tau  = 0}^{LD}  =  - \frac{V}{{B|_{\tau  = 0} }}\left[ {\frac{{\partial ^2 \Phi }}{{\partial V^2 }} + V\frac{{\partial ^3 \Phi }}{{\partial V^3 }}} \right].
\end{eqnarray}

\noindent This term can be obtained from a harmonic lattice dynamics calculation.

\begin{eqnarray}\label{eq8}
B^{\prime}|_{\tau  = 0}^{ZP}  = \frac{V}{{B|_{\tau  = 0} }}\frac{\partial }{{\partial V}}\left( {\frac{{\partial \pi_0}}{{\partial \ln V}}} \right)
\end{eqnarray}

\noindent is the contribution from the zero-point energy. Using Eq.~\ref{eq4}, we estimated this term with the phonon frequencies at different volumes using our Zn(CN)$_2$ potential model.~\cite{Fang 2013} The derivatives were approximated by

\begin{eqnarray}\label{eq9}
\frac{{\partial \pi_0}}{{\partial \ln V}} \approx V_1 \frac{{\pi_0\left( {V_2 } \right) - \pi_0\left( {V_1 } \right)}}{{V_2  - V_1 }}
\end{eqnarray}

\noindent and

\begin{eqnarray}\label{eq10}
&&\frac{\partial }{{\partial V}}\left( {\frac{{\partial \pi_0}}{{\partial \ln V}}} \right)\approx \nonumber \\
&&V_2 \frac{{\pi_0\left( {V_3 } \right) - \pi_0\left( {V_2 } \right)}}{{\left( {V_3  - V_2 } \right)\left( {V_2  - V_1 } \right)}} - V_1 \frac{{\pi_0\left( {V_2 } \right) - \pi_0\left( {V_1 } \right)}}{{\left( {V_2  - V_1 } \right)^2 }}
\end{eqnarray}

\begin{table}[t]
\setlength{\tabcolsep}{4pt}
\caption{The values of cell volume and $\pi_0$ (Eq.~\ref{eq4}) calculated at different pressures using the Zn(CN)$_2$ potential model.~\cite{Fang 2013} These values were further used to calculate $B^{\prime}|_{\tau  = 0}^{ZP}$ using Eqs.~\ref{eq7} and \ref{eq8}.}
\centering
\begin{tabular}{c| c| c| c| c}
  \hline\hline
  $P$ (GPa) & $V$ ({\AA}$^3$) & $\pi_0(\times10^{-3})$ (GPa) & $\partial \pi_0/\partial \ln V$ (GPa)& $B'|_{\tau  = 0}^{ZP}$ \\ [0.5ex]
  \hline
  0.0  &207.2201& 2.0 & 0.12 & $-0.13$ \\ [0.5ex]
  0.2  &206.3314& 1.5 & 0.14 & -- \\ [0.5ex]
  0.4  &205.4938& 0.92 & --  & --  \\ [0.5ex]
  \hline
\end{tabular}
\label{table0}
\end{table}

\noindent where $V_3<V_2<V_1$ are equilibrium volumes at different pressures, and $V_1$ corresponds to $T=0$ and $p=0$. Table~\ref{table0} lists the data from the lattice dynamics of the potential model.~\cite{Fang 2013} The calculated $B^{\prime}|_{\tau  = 0}^{ZP}$ using these data is $-0.13$. Thus, for Zn(CN)$_2$, the contribution from the zero-point energy to the value of $B_0^\prime$ is negative, and the obtained values from the lattice dynamics and MD in the paper should be corrected accordingly, although this correction is too small to have any significant effect on the results.

\begin{acknowledgments}
We gratefully acknowledge financial support from the CISS of Cambridge Overseas Trust (HF), the EPSRC (EP/G004528/2) (ALG) and the ERC (Grant number 279075) (ALG). MD simulations were performed using the CamGrid high-throughput environment of the University of Cambridge. The interatomic potential was developed through our membership of the UK HPC Materials Chemistry Consortium, funded by EPSRC (EP/F067496), using the HECToR national high-performance computing service provided by UoE HPCx Ltd at the University of Edinburgh, Cray Inc and NAG Ltd, and funded by the Office of Science and Technology through EPSRC's High End Computing programme. We thank the STFC for providing access to the ISIS neutron facility.
\end{acknowledgments}

\end{document}